\documentclass[twocolumn]{emulateapj}
\usepackage{color}

\usepackage{times}
\usepackage{graphicx}
\usepackage{amssymb,amsmath}

\shorttitle{Rapid dynamical chaos in an exoplanetary system}
\shortauthors{Deck et al.}

\begin{document}

\title{Rapid dynamical chaos in an exoplanetary system}

\author{Katherine M. Deck\altaffilmark{1}, Matthew J. Holman\altaffilmark{2}, Eric Agol\altaffilmark{3}, Joshua A. Carter\altaffilmark{2,4}, Jack J. Lissauer\altaffilmark{5},  Darin Ragozzine\altaffilmark{2}, Joshua N. Winn\altaffilmark{1}}

\altaffiltext{1}{Department of Physics and Kavli Institute for Astrophysics and Space Research,
Massachusetts Institute of Technology, 77 Massachusetts Ave., Cambridge, MA 02139}
\altaffiltext{2}{Harvard-Smithsonian Center for Astrophysics, 60 Garden St., Cambridge, MA 02138}
\altaffiltext{3}{Department of Astronomy, Box 351580, University of Washington, Seattle, WA 98195}
\altaffiltext{4}{Hubble Fellow}
\altaffiltext{5}{NASA Ames Research Center, Moffet Field, CA 94035}

\begin{abstract}
We report on the long-term dynamical evolution of the two-planet Kepler-36 system, which we studied through numerical integrations of initial conditions that are consistent with observations of the system. The orbits are chaotic with a Lyapunov time of only $\sim$10 years. The chaos is a consequence of a particular set of orbital
resonances, with the inner planet orbiting 34 times for every 29 orbits of the outer planet. The rapidity of the chaos is due to the interaction of the 29:34 resonance with the nearby first order 6:7 resonance, in contrast to the usual case in which secular terms in the Hamiltonian play a dominant role.   
Only one contiguous region of phase space, accounting for $\sim 4.5\%$ of the sample of initial conditions studied, corresponds to planetary orbits that do not show large scale orbital instabilities on the timescale of our integrations ($\sim$ 200 million years). The long-lived subset of the allowed initial conditions are those that satisfy the Hill stability criterion by the largest margin. Any successful theory for the formation of this system will need to account for why its current state is so close to unstable regions of phase space.

\end{abstract}
\section{INTRODUCTION}
Despite the seeming regularity of the Solar system, the planetary orbits are known to be chaotic with a Lyapunov time of $\sim$5 million years \citep{LaskarZ,Wisdom1992}. The hallmark of a chaotic system is sensitive dependence on initial conditions: two trajectories that start arbitrarily close to each other will diverge exponentially on a time scale known as the Lyapunov time. Chaos is seen not only in the orbits of the planets, but also among various satellites and minor bodies of the Solar System \citep{WisdomAsteroid,LecarRev, Gold1}. However, among the hundreds of
multiplanet systems known to exist around other stars, few have orbits measured precisely enough to determine definitively if chaos is present. There is evidence that both the Kepler-11 and 55Cnc systems are chaotic \citep{paper,paper2}, but in the case of 55Cnc, where the masses and inclinations are not well constrained, this conclusion is less certain.

The Kepler$-$36 system consists of a subgiant star of solar mass and the two transiting planets Kepler-36b and c, with orbital periods of 13.8 and 16.2~d and masses of 4.1 and 7.5$M_\oplus$, respectively\citep{CarterandAgol}.  All of the necessary parameters for integration---the bodies' positions and velocities at a reference epoch, as well as their masses---have been measured precisely. This allows study of the true dynamical evolution of the system at a level of detail that has only been achieved for the Solar System and a handful of exoplanetary systems \citep{PulsarPlanets,paper2}. 

   This Letter is organized as follows. Section~\ref{sec:Num} describes our integration methods and the set of initial conditions we use. Our results on the chaotic behavior of the planets, including an explanation of its salient features, are given in Section~\ref{sec:Chaos}. A stability analysis of the system is presented in Section~\ref{sec:Stability}.  We discuss briefly in Section~\ref{sec:Disc} how our results may constrain models of formation of the system.

\section{NUMERICAL METHODS}\label{sec:Num}
The effect on the transit light curves of dynamical interactions between the planets was modeled in a Bayesian fashion using prior knowledge of the host star obtained through astroseismic analysis. This resulted in a posterior joint probability distribution for the bodies' masses and the planetary positions and velocities at a reference epoch  \citep{CarterandAgol}. Initial conditions and masses drawn from this distribution form a representative set because they sample the possible configurations of the planets consistent with the data in a statistically appropriate way. All of the correlations between the uncertainties in these orbital parameters are naturally taken into account.

We studied the dynamical evolution of the Kepler-36 system through numerical integration of $10^4$ initial conditions and masses.\footnote{These initial conditions are published with Carter \& Agol et al. (2012).} Our primary integration scheme was a symplectic n-body mapping \citep{WH1}. We implemented Chambers' symplectic corrector to improve the accuracy of the integrator \citep{Touma,Chambers,LaskarRobutel, W3}, resulting in a relative energy error over the course of our integrations of $\lesssim10^{-12}$. 

``Stepsize chaos" can be a source of error in symplectic integrators \citep{WH2}, but it is negligible when the timestep $\Delta t$ is smaller than the shortest physical timescale in the system by at least a factor of 10 or 20 \citep{Rauch}. Each integration was carried out using a fixed timestep of $\Delta t \le 1\%P_b$.

	Only Newtonian gravitational forces were considered; general relativity is unimportant due to the long timescale for relativistic precession ($\sim 10^8$ days) relative to the secular timescale of $3\times10^4$ days. Tidal effects are discussed in Section~\ref{sec:Disc}. 
\section{RAPID DYNAMICAL CHAOS}\label{sec:Chaos}
A direct way of determining if a particular trajectory is chaotic is to measure how perturbations of that trajectory grow in time. In a chaotic system, a perturbation will grow exponentially in the limit of $t\rightarrow \infty$, with an e-folding time equal to the Lyapunov time. There are as many distinct Lyapunov times as degrees of freedom of the system \citep{Lichtenberg}, but in practice one usually measures the shortest. In addition to a main trajectory, we integrate the tangent equations of the symplectic mapping which govern the behavior of infinitesimally small perturbations of that trajectory.

\begin{figure}
	\begin{center}
	\includegraphics[width=3.7in]{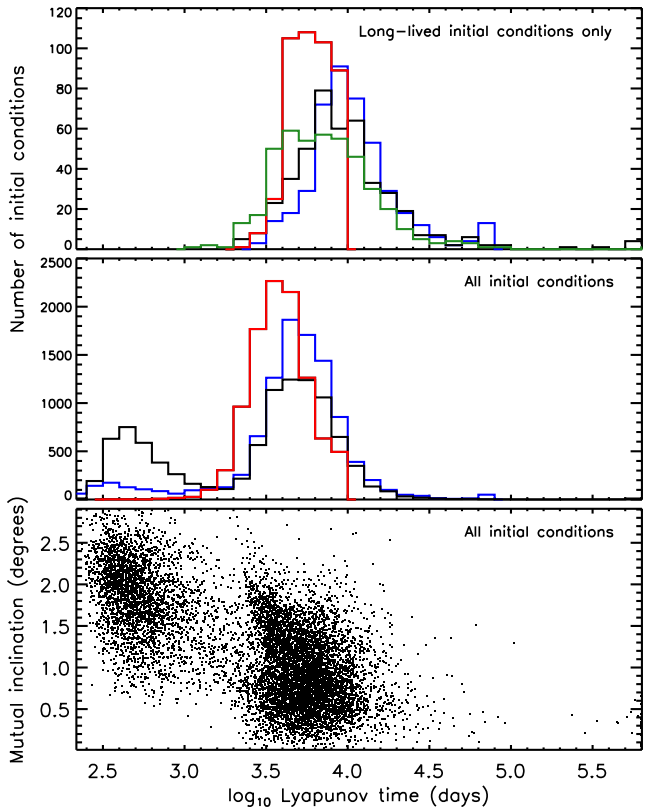}
	\caption{ Distribution of estimated Lyapunov times. {\it Top:} The distribution for the long-lived initial conditions after $5\times 10^7$ days (green), $10^7$ days (black), $10^6$ days (blue), and $10^5$ days (red). {\it Middle:}  The distribution for the entire set of initial conditions, same color scheme.  {\it Bottom:} The trajectories with the fastest chaos correspond to orbits with higher initial mutual inclinations. }
	\label{fig:Lyap_distLong}
	\end{center}
	\end{figure}

The tangent equations were integrated forward by 10$^7$~days for each trajectory. Over 99$\%$ of these initial conditions, including the orbit that best fits the data, are chaotic. The rest do not show exponential divergence on the timescale of the integrations and therefore may be quasiperiodic.  Figure~\ref{fig:Lyap_distLong} shows the distribution of Lyapunov times for the entire set of initial conditions and for those belonging to the long-lived region (where we predict the true orbits must lie, see Section~\ref{sec:Stability}).  	
	The typical Lyapunov time is several thousand days. A more pertinent measure is the ratio of the Lyapunov time to the shortest orbital period. This ratio is $\sim$300 for Kepler-36. In contrast, the Lyapunov time of the Solar System is approximately $2\times 10^7$ orbits of Mercury. One of the most rapid Lyapunov times observed within the Solar System, that of the Saturnian moons Prometheus and Pandora, is 2,000 Promethean orbits \citep{Gold3}. 

We confirmed that similar estimates of the Lyapunov time were obtained when using a smaller timestep ($\Delta t=0.01$~d) and when using a Bulirsch-Stoer integrator \citep{Bulirsch}. As expected, the distribution of Lyapunov times did not change when general relativity was included, which we mimicked using a dipole potential which produces the correct precession rate \citep{Nobili}.

Figure~\ref{fig:Lyap_distLong} shows that the estimated Lyapunov times range over two orders of magnitude, with two clusters centered at approximately 300 and 4,000 days. This was unexpected, since chaotic zones in phase space are typically characterized by a single value of the minimum Lyapunov time \citep{Henon}. Moreover, the estimated Lyapunov times change as the total integration time is increased, with a growing population centered on 300 days.  This suggests that the initial conditions span two nearly disconnected regions of the chaotic zone in phase space, characterized by different estimates of the local Lyapunov time (which is not the absolute shortest Lyapunov time, in the infinite limit), and that trajectories are moving between them. This will be returned to in Section~\ref{sec:Stability}; we turn now to understanding the rapidity of the chaotic behavior.

\subsection{Origin of the chaos}

\begin{figure}
	\begin{center}
	\includegraphics[width=2.6in,angle=90]{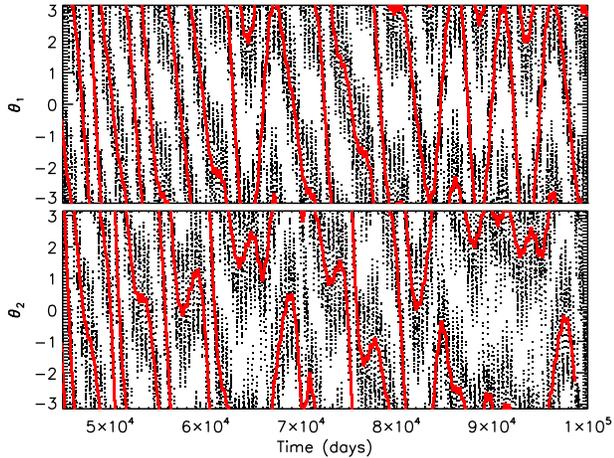} 
	\caption{Chaotic evolution of the resonant angles $\theta_1 = 34\lambda_c-29\lambda_b-5\varpi_b$ and $\theta_2 = 34\lambda_c-29\lambda_b-5\varpi_c$ for a randomly chosen trajectory from the long lived region. The red overlaid points show a smoothed version of the black points to guide the eye. }
	\label{fig:ChaoticAngles}
	\end{center}
	\end{figure}

To gain an understanding of the chaos associated with a Lyapunov time of several thousand days, we sought evidence for resonant behavior.  Mean motion resonances (MMRs) become important when the ratio of the planetary mean motions is close to a rational number, $ n_b/n_c \sim (j+k)/j$, where $j$ and $k$ are mutually prime integers. Numerically we find that the angles characterizing the 29:34 eccentricity-type MMR, $\theta_1 = 34\lambda_c-29 \lambda_b-5\varpi_b$ and $\theta_2= 34\lambda_c-29 \lambda_b-5\varpi_c$ (and the linear combinations of $\theta_1$ and $\theta_2$ appearing in the interaction Hamiltonian), to be behaving chaotically. Here $\lambda$ and $\varpi$ refer to the mean longitude and longitude of periastron of the orbits. Chaotic alternations of these angles between circulation and libration are shown for a particular trajectory in Figure~\ref{fig:ChaoticAngles}. 				  It may be surprising that such a high order (large $k$) MMR would be important, as resonance widths in phase space scale as $k$ factors of eccentricities and inclinations (and therefore are usually negligible for nearly circular and coplanar orbits). However, the Laplace coefficients, which also factor into the widths, are large when the semimajor axis ratio of the planets is near unity, as is the case for Kepler-36. This allows the 29:34 resonance to be important. 

To understand why the 29:34 resonant angles should behave chaotically, we appeal to the resonance overlap criterion  \citep{Chirikov}. This states that chaos will ensue if the angles $\theta_2$ and $\theta_1$, when neglecting the interaction between them, are both analytically calculated to be librating in the same region of phase space. This can occur if the resonance widths, which are functions of the eccentricities, become large enough or if the resonance centers, determined to be where $\dot{\theta}_1=0$ and $\dot{\theta}_2=0$, coincide. We find that the vast majority of the initial conditions exhibit oscillations with a period of several thousand days in the angle $\varpi_b-\varpi_c$, indicating that $\dot{\varpi}_b \sim \dot{\varpi}_c$, i.e. that the resonant islands are overlapping.

The approximate equations of motion for resonant angles resemble those of coupled pendulums driven by the oscillations in the eccentricities and the periastra of the two planets. This correspondence can be used to show that the resulting chaotic behavior should have a Lyapunov time similar to the period of these oscillations \citep{MH1}. For Kepler-36 this period is several thousand days, similar to the Lyapunov time, supporting this explanation of the origin of the chaos.

This explanation alone is not surprising, as the connection between chaotic behavior and MMR overlap is well known. Why then is the Lyapunov time (relative to the smallest orbital period) so short for this sytem?  In previously known examples MMR overlap, the periodic driving of the resonant angles by the eccentricity and periastra are caused by secular effects. In the Kepler-36 system, the driving is dominated by the effects of a nearby first-order MMR ($P_b/P_c \sim 6/7$), which appear at a lower order (in the eccentricities) in the interaction potential.  It can be shown that the proximity to the 6:7 resonance does not alter the qualitative character of the standard secular solution for the eccentricities and periastra but that it does affect the frequencies of the oscillations significantly. Following Malhotra et al. (1989), we estimate the shorter of the two modified secular timescales to be $\sim8,000$ days. We confirmed that the 6:7 terms are crucial for producing the correct behavior of the eccentricities and periastra numerically as well. 

The 6:7 inclination-type terms appear in the Hamiltonian at the same order in the inclinations as the secular terms and hence should not be as crucial as the 6:7 eccentricity terms.  Trajectories that have been projected onto the invariable plane remain chaotic with a Lyapunov time of several thousand days, confirming that a coplanar model is sufficient to explain the chaos.

\begin{figure}
	\begin{center}
	\includegraphics[width=3.4in]{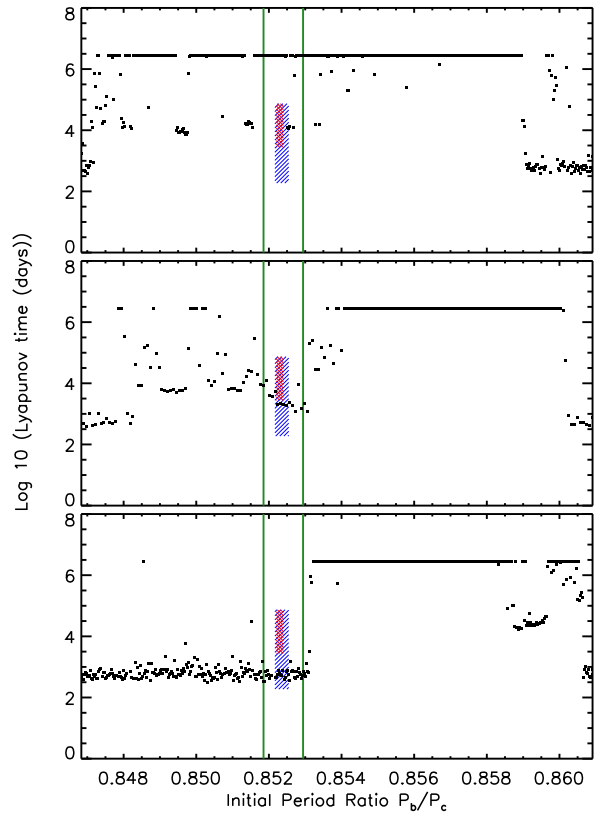}
	\caption{Structure of the phase space near the Kepler-36 initial conditions.  Red indicates the span of the long-lived trajectories on this plane and blue the entire sample. Black points correspond to trajectories that do not fit the data. Green lines indicate the nominal location of the 23:27 and 34:39 MMR. Only $a_b$ varies within each panel. {\it Top:}  $(e_b,e_c) \sim (0.02,0.0)$. {\it Middle:}  $(e_b,e_c) \sim (0.021,0.006)$. {\it Bottom:}  $(e_b,e_c) \sim (0.036,0.0)$.  }
	\label{fig:Scan}
	\end{center}
	\end{figure}

A much smaller subset of the initial conditions shows chaotic behavior of the angles characterizing the 23:27 eccentricity-type MMR. Figure~\ref{fig:Scan} delineates the chaotic zones (Lyapunov time $\lesssim 10^6$~d) associated with separate MMR, for different eccentricities, demonstrating that both the 23:27 and 29:34 resonances can be important. Moreover, the chaotic region is not always confined to a small range in period ratio, indicating that the orbits of the planets may not always be stable, or constrained.

\section{STABILITY OF THE KEPLER-36 SYSTEM}\label{sec:Stability}
There are two primary modes of instability in planetary systems. First, planets may have close encounters or collisions. An analytic criterion is available for two-planet systems, based on conservation of angular momentum $L$ and energy $E$, which can exclude this possibility \citep{Marchal261982}, in which case the planets are labeled Hill stable. The (sufficient but not necessary) criterion for stability is
	\begin{equation}\label{Hill}
	h \equiv -c L^2 E > h_{crit}
	\end{equation}
	where $h_{crit} = 1+3^{4/3}m_b m_c m_\ast^{-2/3}/(m_b+m_c)^{4/3}+...$ and $c = 2(m_\ast+m_b+m_b)/(G^2 (m_b m_c+m_c m_\ast +m_b m_\ast)^3)$. Here $m_\ast$ refers to the mass of the star.

Even if the planets never have close encounters, repeated weak interactions can lead to a second type of instability, in which the gradual exchange of angular momentum and energy between the planets results in drastic orbital variations. This is known as Lagrange instability. There is no known analytic criterion for Lagrange stability, but numerical integrations can demonstrate that a given trajectory does not show unstable behavior on the timescale of the integration and therefore can be considered to be ``long-lived".

A preliminary stability analysis of the Kepler-36 system determined that $\sim 9\%$ of the initial conditions did not satisfy the Hill stability criterion \citep{CarterandAgol}.  It was left to future work to determine whether the initial conditions were Lagrange long-lived.

\subsection{Lagrange stability analysis}

We integrated all $10^4$ trajectories for $2.5 \times 10^9$~d and found that while only $30\%$ showed Lagrange unstable behavior during the first $\sim 10^8$~d, this percentage had increased to $\sim75\%$ by the end of the integrations. We classified Lagrange instability as variations in semimajor axis, eccentricity, and inclination that were different from their running average by greater than $10\%$. In general, the trajectories that were Lagrange unstable and those that did not show instability during these integrations were well mixed, but we could identify a single contiguous region in phase space, accounting for $\sim 4.5\%$ of the initial conditions, which had no unstable trajectories on these timescales. This region is characterized by lower eccentricities and inclinations, shows no systematic difference in the goodness-of-fit to the data, and contains five of the candidate quasiperiodic trajectories. One hundred randomly chosen initial conditions from this ``long-lived core" were integrated for 200 million years ($> 5 \times 10^9 P_b$). Of these, only four exhibited instability; all came from the borders of the long-lived core. 
  
\begin{figure}
	\begin{center}
	\includegraphics[width=3.5in]{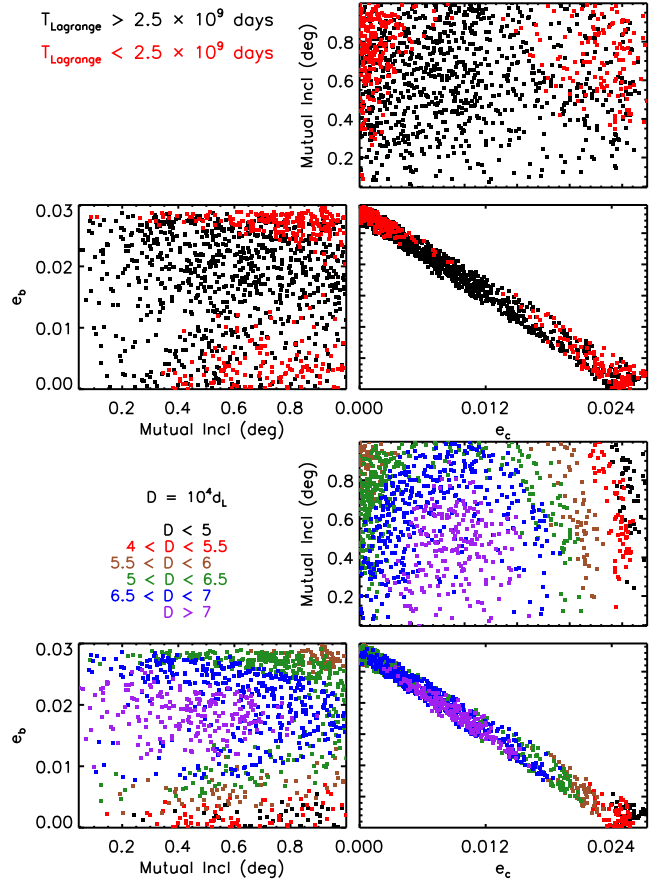}
	\caption{ Relationship between Hill and Lagrange stability. {\it Upper Plots:} The region of orbital element space where orbits can be long-lived. Lagrange unstable orbits are shown in red, orbits that have not undergone Lagrange stability in 2.5 $\times 10^9$ days are in black.  {\it Lower Plots:} By grouping initial conditions based on their values of $d_L$ we identify the Lagrange long-lived core.}
	\label{fig:HillLag}
	\end{center}
	\end{figure}

	The initial conditions belonging to the long-lived core satisfy
	\begin{equation}\label{Lagrange}
	h> h_{crit}+\epsilon
	\end{equation}
		where the value of $\epsilon \sim 0.0007$ is determined numerically.  In other words, orbits that are Lagrange long-lived must satisfy a criterion nearly identical to the Hill criterion, but with a slightly different critical value. This same relationship was found for Jupiter-mass planets \citep{BG1}. Our work suggests that the same link between Hill and Lagrange stability applies across orders of magnitude in the parameter $m_{planet}/m_{star}$, though the value of $\epsilon$ depends on this parameter. Contours of constant $d_L \equiv h - h_{crit}$ are indicated in Figure~\ref{fig:HillLag}.
		
		 The initial conditions belonging to the long-lived region are those with the smallest values of the angular momentum deficit (AMD), a quantity which parametrizes how far the orbits are from purely circular and coplanar. It is straightforward to show that $d_L$ is a function of the AMD for nearly circular, coplanar orbits, explaining the observed dependence.

		To explore the connection between $d_L$ and Lagrange instability, we focused on a group of 475 Hill stable trajectories with a uniform distribution in $d_L$.  There is a sharp transition at $d_L = 0.0007$, where the trajectories change from being mostly Lagrange unstable to being entirely Lagrange long-lived. A smaller value of $d_L$ correlates with a shorter time to show Lagrange instability and with smaller minimum approach distances between the two planets.
		
		  It is not obvious what longer integrations would reveal regarding the trajectories that appear to be long-lived on the timescale of our integrations but do not satisfy equation~\eqref{Lagrange} (i.e. are not in the long-lived core). Previous work indicates that unless these these orbits are protected by some resonance mechanism they too will show Lagrange instability on relatively short timescales \citep{Gladman,BGRes}. We predict that future observations will show that the true orbits of the Kepler$-$36 planets belong to the long-lived core.
 
  By restricting the initial conditions to the long-lived core we refine the system parameters. This results in median values and $84.2\%$ and $15.8\%$ confidence intervals for the masses and radii of the planets of $m_b = 4.32 M_\oplus  \mbox{}^{+0.19} _{-0.20}, m_c = 7.84 M_\oplus \mbox{}^{+0.33} _{-0.36}, R_b = 1.49 R_\oplus  \mbox{}^{+0.035} _{-0.035}$, and $R_c= 3.68 R_\oplus \mbox{}^{+0.056} _{-0.055}$. The best fit values are $m_b = 4.11 M_\oplus, m_c = 7.46 M_\oplus, R_b = 1.46 R_\oplus,$ and $R_c  = 3.59 R_\oplus$. The mutual inclination is constrained to be less than one degree.

		   Although these planets seem to be on the brink of instability, the orbits are stable to small perturbations. In particular, we confirmed that there are hypothetical third planets (with a period of $2.6P_c$ and $M\le 5M_\oplus$) that do not disrupt the two planet system on seven million year timescales.
		 
\subsection{Transition to Lagrange instability}\label{sec:ChaoticDiff}	
\begin{figure}
	\begin{center}
	\includegraphics[width=3.3in]{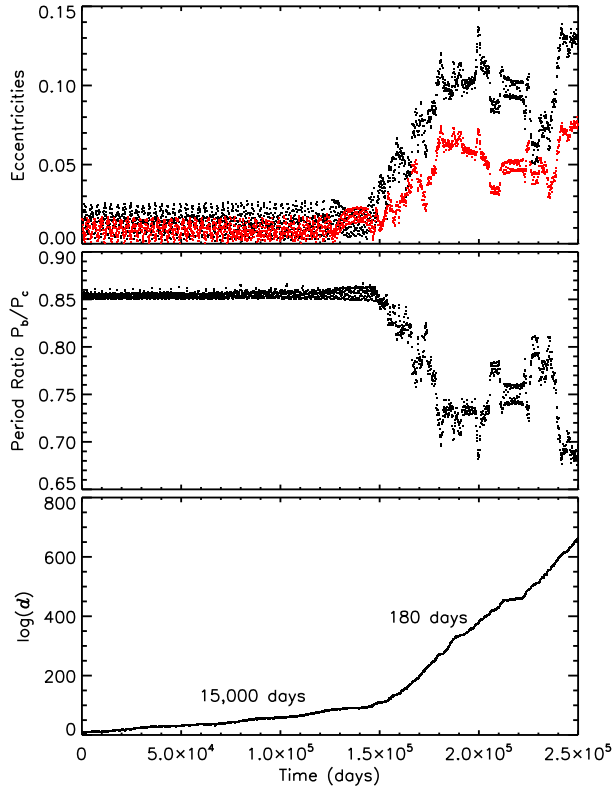}
	\caption{Period ratio and eccentricity evolution ($e_b$ in black, $e_c$ in red) for a Lagrange unstable orbit. When the orbit begins to change qualitatively in an erratic way, the slope of $\log(d)$ changes, indicating a change in the estimated Lyapunov time. $\log(d)$ is the natural logarithm of the norm of distance between two initially nearby trajectories in phase space. }
	\label{fig:LagrangeFast}
	\end{center}
	\end{figure}

We now return to the evolution of the distribution of Lyapunov times shown in Figure~\ref{fig:Lyap_distLong}. We find that the onset of Lagrange unstable behavior occurs when the trajectory moves between the two peaks of the distribution (at $\sim300$ and $\sim$4,000 days). A typical Lagrange unstable trajectory is shown in Figure~\ref{fig:LagrangeFast}. 
				 The eccentricities and semimajor axes remain nearly constant for about $10^5$ days, after which the orbital elements vary dramatically. At the same time, the estimated Lyapunov time changes from roughly 15,000 days to 180 days.

The ability for trajectories to cross between the two nearly disconnected regions of the chaotic zone (which are characterized by different estimates of the local Lyapunov time and also are chaotic for different reasons) can be understood as a consequence of resonance overlap \citep{Mardling,MH2}. As Figure~\ref{fig:Scan} shows, separate MMRs merge at higher eccentricities. Our hypothesis is that chaotic diffusion of eccentricities to higher values results in MMR overlap, at which point a bottleneck between the two regions is formed. Once this occurs, trajectories can enter the Lagrange unstable region where they explore a broader range of period ratios.  This chaotic diffusion may occur for all of the chaotic trajectories, in which case the distribution of Lyapunov times will approach an asymptotic form with all trajectories belonging to the 300 day peak as longer integrations are performed.

\section{DISCUSSION}\label{sec:Disc}
We have presented a dynamical analysis of the Kepler-36 system which has yielded several surprising results. The orbits are chaotic with an extremely short Lyapunov time, yet some still manage to be long-lived. The closeness of this system to instability is an intriguing feature of Kepler-36. In particular, did the planets form with orbits contained in the long-lived core? Alternatively, did some dissipative process drive them into this long-lived configuration? 

It would seem that tidal dissipation is negligible for this system, despite the proximity of the planets to the star. If tidal dissipation were important then the planets would have been on more eccentric orbits in the past, and, assuming the orbital separation was not very different at that time, it is unlikely such a configuration would have been stable.

Alternatively, the planets could have been closer together if they were protected by a resonance. A commonly discussed mechanism for forming compact multiplanet systems is convergent migration in the gaseous protoplanetary disk or the disk of remnant planetesimals \citep{Papaloizou}. Perhaps the Kepler-36 planets formed at large orbital distances, and migrated until being locked into a 6:7 resonance. After migration ended, subsequent tidal evolution could have driven the planets apart to their current configuration. However, we recognize that it may be challenging to create closely packed resonant systems through conventional migration mechanisms \citep{Rein}, though other studies find that it is possible to produce systems like Kepler-36 \citep{IdaLin,Ogihara,Pierens}. Therefore this system may contain important clues about the relevance of convergent migration to the formation and dynamical evolution of planetary systems.

\acknowledgments
We thank J. Wisdom for sharing with us his code that calculates coefficients in the disturbing function, as well as other members of the Kepler TTV team for helpful comments on the text.  EA acknowledges support for this work was provided by NSF Career grant AST-0645416.

\clearpage

\end{document}